\documentclass[a4,12pt]{article}
\usepackage{amssymb}
\usepackage{pifont}
\usepackage{amsthm}
\usepackage{colortbl}
\usepackage{rotating}
\usepackage{makecell}
\usepackage{natbib}
\usepackage[dvipsnames]{xcolor}
\usepackage{booktabs}
\usepackage{booktabs}
\usepackage{afterpage}
\usepackage[normalem]{ulem}
\usepackage{amsmath}
\usepackage{hyperref}
\hypersetup{
	colorlinks = false,
	hidelinks = true
}
\usepackage[toc,page]{appendix}
\usepackage{amsthm}          

\theoremstyle{plain}

\theoremstyle{definition}

\theoremstyle{remark}

\title{Artificial intelligence and financial crises\thanks{Corresponding author Jon Danielsson, J.Danielsson@lse.ac.uk. Updated versions of this paper can be downloaded from from \url{modelsandrisk.org/appendix/AI}. We thank the
editor and two anonymous referees for helpful comments as well as the Economic and Social Research Council (UK), grant numbers ES/K002309/1, ES/R009724/1 and ES/Y010612/1 for their support. Any opinions and conclusions expressed herein are those of the authors and do not necessarily represent the views of the Bank of Canada.}
}

\author{Jon Danielsson\\\textit{London School of Economics} \and Andreas Uthemann\\\textit{Bank of Canada}\\\textit{Systemic Risk Centre, London School of Economics}}

\date{July 2025\\Forthcoming in the Journal of Financial Stability}

\begin{document}
\maketitle
\thispagestyle{empty}

\begin{abstract}
\noindent
The rapid adoption of artificial intelligence (AI) poses new and poorly understood threats to financial stability. We use a game-theoretic model to analyse the stability impact of AI, finding that it amplifies existing financial system vulnerabilities --- leverage, liquidity stress and opacity --- through superior information processing, common data, speed and strategic complementarities. The consequence is crises become faster and more severe, where the likelihood of a crisis is directly affected by how effectively the authorities engage with AI. In response, we propose that the financial authorities develop their own AI systems and expertise, establish direct AI-to-AI communication, implement automated crisis facilities and monitor AI use.
\end{abstract}

\newpage

\clearpage

\section{Introduction}

The relationship between artificial intelligence (AI)\footnote{\cite{Russel2019} and \cite{RussellNorvig2010} are particularly useful for a general overview of AI.} and financial crises is poorly understood, motivating our work. We use a game-theoretic framework to identify how AI impacts financial stability, finding that it can both stabilise and destabilise the financial system, depending on the severity of the shocks it faces and how prepared the financial authorities are. As current regulatory frameworks cannot effectively prevent and resolve such AI-induced crises, we outline how the financial authorities can adapt to AI-driven risks through upgraded supervisory tools, proactive intervention mechanisms and institutional readiness.

While there is no universal definition of AI, the notion of it as a rational maximising agent\footnote{One of the definitions given by \cite{RussellNorvig2010}.} is particularly useful for the analysis of financial stability since it resonates with methodological approaches taken by most macroprudential investigators. That implies standard frameworks for analysing financial stability are well-suited for studying the impact of AI provided they consider AI's speed and the particular strategic approaches it applies to tasks given to it. To that end, we adopt a global games model \citep{morris1998unique} and highlight three key mechanisms through which AI transforms market dynamics: superior information processing that makes signals more precise, common information sources that facilitate coordination and speed advantages that give rise to preemption motives. 

Our model yields specific predictions about AI's impact on crisis dynamics. First, AI makes transitions between stable and crisis states more abrupt. Second, AI increases the likelihood of multiple equilibria and coordinated runs due to the fundamental factors that have driven financial crises for centuries, ever since the first modern one in 1763 \citep{SchnabelShin04, DanielssonBook2020}: Excessive leverage, liquidity preference during stress and system opacity.

When we apply these theoretical predictions to a more practical institutional structure of the financial system, we find four key areas of concern. First, the malicious use of AI --- whether to find loopholes, manipulate markets, orchestrate terrorist attacks on financial infrastructure, or enable nation-states to engage in unconventional warfare --- poses serious threats to stability. Even a small, coordinated misinformation shock can push collective beliefs beyond the tipping point and trigger a self-fulfilling crisis.

Second, wrong-way risk, which refers to a situation in which the volume of risky activities increases in line with the riskiness of those activities. AI gives rise to particularly dangerous forms of wrong-way risk because of how it gains trust, implying that AI reliability is the lowest precisely when it is needed for the most important decisions --- the AI wrong-way risk. This may occur when margins set by AI are downward biased, in turn encouraging risky positions and their associated risks. 

Third, there is the potential for synchronised behaviour --- a consequence of the complementarity of most consequential decisions, where the optimal move of one market participant incentivises others to do the same. This correlated behaviour can be especially damaging during crises, amplifying both market booms and busts. In other words, AI is procyclical. As signal correlation rises, the stabilising heterogeneity of equilibrium behaviour vanishes, increasing the risk of procyclical herd behaviour.

Finally, the increased use of AI affects the speed, intensity, and frequency of crises. When faced with a shock, an AI engine --- like a human decision-maker --- has a range of options it can deploy in response. But they all boil down to two fundamental choices. Run or stay. Stabilise or destabilise. The consequences of getting it wrong can be significant, including substantial losses and even bankruptcy. AI's superior ability to both quickly analyse a complex evolving situation and make quick, firm decisions implies it will make crises particularly quick and vicious. When it acts as a crisis amplifier, what might have taken days or weeks to unfold can now happen in minutes or hours. A greater price impact, combined with shorter reaction times, shortens the adjustment window, shifting the crisis threshold into a state space that was previously considered safe.

The financial authorities --- central banks and regulators tasked with maintaining system stability --- find it challenging to effectively address the risks arising from these four areas of concern. The problem for the authorities is that they are already losing an arms race with the private sector. 

To remain effective, we propose that the authorities focus on several key areas: developing their own AI engines and expertise, establishing AI-to-AI links, implementing automatic facilities, public-private partnerships and particular monitoring of private sector AI use. While most authorities appear to prefer a slow, deliberative and conservative approach to AI, several have now recognised AI's implications for financial stability; see recent publications from the IMF \citep{IMF2024ai}, the BIS \citep{BISAI2024q, BISAI2024po}, the FSB \citep{FSB2024aireport}, the ECB \citep{LeitnerSinghKraaijZsamboki2024}, and the Bank of England \citep{BoEMonitoring2025}.

The organisation of the paper is as follows. After the introduction, we develop our game-theoretic framework for analysing AI's impact on financial stability in Section \ref{sec:model}. Section \ref{sec:channels} examines the four key ways AI can undermine financial stability. We propose specific policy responses in Section \ref{sec:policy}. Section \ref{sec:conclusion} concludes.

\section{A Game-Theoretic Framework for AI and Financial Stability}
\label{sec:model}

To understand how AI affects financial stability, we develop a formal framework that captures both the unique characteristics of AI and its implications for coordination in financial markets. The key to understanding financial crises lies in how financial institutions optimise --- they aim to maximise profits given acceptable risk. When translated into operational behaviour, Roy's (1952) criterion \nocite{Roy52} is particularly useful — stated succinctly, maximising profits subject to not going bankrupt. When financial institutions prioritise survival, their behaviour changes rapidly and drastically. They hoard liquidity and choose the most secure and liquid assets, such as central bank reserves. This leads to bank runs, fire sales, credit crunches and all the other undesirable behaviours associated with crises.

Strategic complementarities play a key role in the rapid, strong shift in behaviour from short-term profit maximisation to survival. When financial institutions find their counterparties are deciding on how to react to a shock, they know the actions of one firm adversely affects the others, making it paramount to react as fast as possible. It is optimal to preempt, that is, to act early and strongly.

Three characteristics of AI engines particularly impact how AI affects stability. First, AI systems process data more efficiently than humans, likely giving them more precise signals about market fundamentals and reducing uncertainty. Second, AI systems often rely on similar data, methodologies and especially model architectures. Third, AI can execute transactions much faster than human traders.

The interaction of these mechanisms yields specific predictions about how the deployment of AI affects financial stability. More precise signals reduce the grey area where fundamentals alone do not determine individual market participants' actions, making aggregate behaviour more binary: either no one ``runs'' (e.g., sells assets or withdraws funding) or everyone does, with transitions becoming sharper and more sudden as fundamentals deteriorate. Common information sources facilitate coordination among AI agents through implicit learning rather than explicit communication, making multiple equilibria more likely, particularly in markets with high AI penetration. Speed advantages create strong preemption motives, with the presence of slower human agents increasing the likelihood of AI-initiated runs. Strategic complementarities amplify these effects through contagion to human agents.

To systematically analyse these dynamics, we adapt the global game's framework of \cite{morris1998unique}, as it is particularly suitable for analysing coordination problems with strategic uncertainties, where agents make decisions without knowing others' actions. This framework allows us to develop theoretical predictions about how a putative AI crisis might unfold.

\subsection{Setup}

We adapt \nocite{SakovicsSteiner2012} Sakovics and Steiner's (2012) model setup to a financial market with a unit mass of investors (AI and human) who must decide whether to ``run'' or ``stay'' when faced with a shock that might lead to a crisis. 

There is a continuum of investors of unit mass indexed by $i \in [0,1]$. A fraction $\mu$ of investors are AI while the rest are human. The fundamental state $\theta \in [\underline{\theta}, \bar{\theta}]$ represents the underlying strength of the financial system, with higher values indicating greater resilience. Let $a \in [0, 1]$ denote the mass of investors who decide to run, where, in particular, a crisis occurs if the mass of investors running exceeds the system's resilience: $a \geq \theta$.

An investor's payoff from choosing to stay given $a$ and $\theta$ is:
\begin{equation}
 u(a, \theta) =
 \begin{cases}
 b - c & \text{if } a \leq \theta \text{ (no crisis)},\\
 -c & \text{if } a > \theta \text{ (crisis)},
 \end{cases}
\end{equation}
where $c$ is the cost of staying (e.g., the cost of financing a position), and $b > c$ is the payoff if no crisis occurs. The payoff from running is normalised to $0$ and does not depend on $\theta$ or $a$.

Table \ref{tab:payoffs} summarizes this payoff structure:

\begin{table}[h!]
 \centering
 \begin{tabular}{*{3}{c}}
 \toprule
 & crisis $(a \geq \theta)$ & no crisis $(a < \theta)$ \\
 \midrule
 run & $0$ & $0$ \\
 stay & $-c$ & $b - c$ \\
 \bottomrule
 \end{tabular}
 \caption{Payoffs for run and stay depend on $a$ and $\theta$.}
 \label{tab:payoffs}
\end{table}

An investor faces two types of uncertainty when deciding on the optimal strategy: fundamental uncertainty about how vulnerable the system is to runs ($\theta$) and strategic uncertainty about how other market participants will act ($a$).

\subsection{Information Structure and AI Mechanisms}

Each investor receives two noisy signals about the fundamental state $\theta$: a private signal $x_i = \theta + \sigma\eta_i$ and a public signal $y = \theta + \sigma_p u$. All investors observe $y$, but only investor $i$ observes $x_i$. The common noise term $u$ and the private noise terms $\eta_i$ for all investors $i \in [0,1]$ are random variables with continuous densities and support $[-1/2, 1/2]$ and are mutually independent. $1/\sigma^2$ is the precision of the private signal and $1/\sigma_p^2$ is the precision of the public signal. A higher precision implies a more accurate signal about the fundamental. Investors' prior about $\theta$ is the uniform distribution on $[\underline{\theta}, \bar{\theta}]$ with $\underline{\theta} < \sigma/2$ and $\bar{\theta} > \sigma/2$. 

\subsubsection{Mechanism 1: Superior Information Processing}

Suppose for the moment that there are no public signals. In equilibrium, investors follow threshold strategies, running if and only if their private signal falls below a threshold $x^*$:
\begin{equation}
a_i(x_i) = 
\begin{cases}
 0 & \text{if } x_i > x^*,\\
 1 & \text{if } x_i \leq x^* .
\end{cases}
\end{equation}

When all investors act according to these threshold strategies, a run occurs whenever $\theta < \theta^{\star}$, where $\theta^{\star}$ is the unique value satisfying\footnote{At equilibrium, the investor with signal $x^* = \theta^*$ is indifferent between ``stay'' and ``run'' because each investor best-responds to a uniform belief about their opponents' action choices \citep[see]{morris2003global}.}
\begin{equation}
\int_0^1 u(a, \theta^{\star})da = 0 \; \Rightarrow \; \theta^{\star} = \frac{c}{b} .
\end{equation}

AI systems typically process information more efficiently than humans,\footnote{Recent evidence suggests that AI systems can achieve superior signal precision in predicting financial crises \citep{Fouliard2021, Hellwig2021, Batsuuri2024}, though this advantage may not extend to unprecedented crisis events with unique characteristics.
} resulting in more precise private signals (lower $\sigma$). While this does not affect the probability of a successful run --- $\theta^{\star}$ does not depend on $\sigma$ --- it affects equilibrium behaviour by reducing uncertainty about the fundamental state, making the critical threshold $\theta^{\star}$ for a crisis more sharply defined in terms of the aggregate action. As the precision of the private signal increases (as $\sigma \to 0$), all investors coordinate on stay when $\theta \geq \theta^{\star}$ and on run when $\theta < \theta^{\star}$. This leads to fewer ``unnecessary'' individual run decisions when $\theta > \theta^{\star}$ but increases the completeness of runs when $\theta < \theta^{\star}$.

\subsubsection{Mechanism 2: Common Information Sources}

When investors observe both private and public signals, they form beliefs about the fundamentals by combining these information sources. \cite{hellwig2002public} shows that when public information is sufficiently precise relative to private information, multiple equilibria can emerge. This happens because the public signal $y$ serves as a coordination device, allowing investors to synchronise their actions.

AI systems are particularly susceptible to this synchronisation effect because they draw on common training data and methodologies, share similar model architectures and objective functions and process similar market data in comparable ways. These factors enhance the public component of information in AI systems, thereby increasing the weight placed on public signals relative to private signals. Formally, this can be represented as AI agents having a higher relative precision of public to private information ($\sigma/\sigma_p$).

The tendency toward common information sources in AI systems creates a potential for coordinated behaviour, similar to the portfolio insurance mechanisms analysed by \citet{gennotteleland90}. This coordination can occur without explicit collusion through shared data sources and similar training approaches. \citet{CalvanoEtAl2020} provide an example of a market game where Q-learning algorithms engage in collusive pricing using commonly observed price data.

\subsubsection{Mechanism 3: Speed Advantage and Preemption}

To capture AI's speed advantage, we introduce heterogeneity between AI and human agents. Suppose the population consists of two groups: human agents (group $H$) of measure $1-\mu$, and AI agents (group $A$) of measure $\mu$. The key difference is that AI agents have an advantage over humans when deciding to run --- they can execute transactions more quickly and efficiently, avoiding the execution delays and price impact that humans might face.

While humans succeed with probability $p < 1$ when choosing to run, AI agents succeed with probability 1. An unsuccessful run yields the same payoff as a stay.

Following \cite{SakovicsSteiner2012}, we can derive the equilibrium threshold $\theta^{\star}$ in this heterogeneous setup:\footnote{For a detailed derivation, see Appendix \ref{app:ai-crises-theory}.}
\begin{equation}
\theta^{\star} = \frac{c}{b}[(1-\mu) p + \mu] .
\end{equation}

Comparing this to a humans-only market ($\theta^{\star} = p \cdot c/b$) reveals that the presence of AI increases the critical threshold, making coordination on the run equilibrium more likely. This occurs because AI agents can withdraw liquidity more efficiently, imposing larger externalities on other market participants by increasing their incentive to withdraw early. This mechanism demonstrates how AI can make financial markets more fragile even without changes in preferences or information quality.

The preemption dynamics created by AI's speed advantage parallel those analysed by \citet{AbreuBrunnermeier03} in the context of bubbles and crashes. However, AI systems can act much more quickly than human traders, potentially compressing the timeframe of market adjustments from days to minutes, as seen in some market events \citep{kirilenko2017flash}.

\section{Channels of AI Impact on Financial Crises}
\label{sec:channels}

Having established a theoretical framework for how AI affects financial stability, we now examine specific vulnerability channels in real-world financial markets. The three theoretical mechanisms --- superior information processing, common information sources, and speed advantage --- interact with extant economic vulnerabilities and institutional structure to create four distinct channels of AI-induced financial risk: malicious use, wrong-way risk, synchronisation and speed.

\subsection{Malicious Use}\label{sec:channels_malicious}

Technology has always both helped improve the financial system and provided new ways to destabilise it. For instance, the telegram in 1858 was quickly employed to transmit stock prices across the Atlantic, while algorithmic trading has been implicated in flash crashes \citep{kirilenko2017flash}. AI is the same but with important qualitative differences that relate to the three mechanisms described in our framework. The model in the previous section suggests that even a small, coordinated misinformation shock can push collective beliefs past the run threshold and trigger a self-fulfilling crisis, similar to the political manipulation strategies analysed in \cite{edmond2013information}.

Malicious use of AI in financial markets can occur through several channels. First, bad-faith actors can exploit AI's superior information processing capabilities to identify regulatory loopholes or engage in market manipulation. Terrorist groups could leverage AI's coordination capabilities and speed to orchestrate synchronised attacks on financial infrastructure. Of particular concern is nation-states recruiting AI for unconventional warfare to target vulnerabilities in adversaries' financial systems while maintaining plausible deniability.

These threats pose a particular challenge to those tasked with defending the financial system --- what we term the ``defender's dilemma''. The attackers need only to find a single vulnerability, whereas the defenders must monitor the entire system against potential attacks. The attackers consequently require considerably fewer resources than the defenders, an asymmetry that gets worse with the increasing ability of AI systems.

As a rational maximising agent, AI can also act maliciously without explicit human direction. \citet{scheurer2023technical} provide a telling example: a large language model (LLM) was instructed to comply with securities laws and maximise profits. The aim was to investigate how AI would address these two potentially conflicting objectives. When given private information, the LLM engaged in illegal insider trading and then lied to its human overseers. This demonstrates a form of model risk that emerges from providing AI with agency. It might find unexpected and potentially harmful paths to attain the goals it is given.

AI's speed advantage makes it difficult to guard against such behaviour. A potential solution is to maintain humans in-the-loop to make decisions based on AI recommendations, or humans on-the-loop to supervise AI. However, human experts cannot effectively oversee AI operating at machine speeds, and in competitive markets, those who remove humans from the loop can gain the upper hand.

The existing regulatory structure struggles with AI-specific principal-agent problems. Conventional incentive mechanisms which work well with human agents, such as rewards and punishments, lose traction when the agent is an optimisation algorithm that cares only about its objective function \citep{Amodei2016,Leike2018}. Mis-specified goals invite reward hacking and other forms of specification gaming \citep{Krakovna2020}, so the classic one-sided principal-agent problem becomes a two-sided principal-agent-AI problem in which supervisors must monitor both the human intermediaries and the opaque learning system they control.

\subsection{Wrong-Way Risk}\label{sec:channels_wrongway}

Wrong-way risk refers to situations where exposure to risk increases when that risk is most severe. AI introduces particularly dangerous forms of wrong-way risk, stemming from how it gains trust. AI builds trust incrementally by excelling at simple, repetitive tasks, leveraging its information processing advantages. As AI systems demonstrate competence in increasingly complex tasks, financial institutions progressively entrust them with more consequential decisions. This may culminate in an AI version of the Peter principle, where AI systems are promoted to increasingly critical roles until their capabilities no longer match the job requirements. 

The typical response to such risks is to maintain that AI will not be used for important decisions, ensuring humans will always remain in/on-the-loop, except for the most innocuous tasks. However, such guarantees face credibility problems in the competitive financial sector, where AI's speed advantages provide significant competitive benefits, as discussed above.

The most serious wrong-way risk arises from a fundamental mismatch between AI's information processing strengths and the nature of financial crises. AI excels at pattern recognition based on historical data but struggles with unprecedented scenarios --- the unknown unknowns that characterise financial crises. As \citet{DanielssonBook2020} argues, every financial crisis is unique in its specifics. While the same three fundamental vulnerabilities --- excessive leverage, liquidity preference, and system opacity --- cause all crises, the precise mechanisms differ significantly from one crisis to the next. Consequently, while AI can be a considerable aid in identifying stress and analysing crises ex-post \citep{Hellwig2021,Fouliard2021,Batsuuri2024}, it will not be very helpful in predicting and preventing crises.

The problem is exacerbated by the fact that we cannot predict how future financial institutions, political leaders and regulators will react to stress. Their reaction functions remain unknown because policymakers like to keep strategic ambiguity, while rotating senior decision-makers. The legal and political environment will be different, and they do not even know how they will react without knowing the concrete circumstances. Consequently, these reaction functions cannot be captured in historical datasets. Something that is not in a dataset cannot be learned by neural networks, no matter how sophisticated.

This creates a particularly dangerous form of wrong-way risk where AI's decision quality is inversely proportional to the importance of the decision at hand. In normal times, when decisions have relatively minor consequences, AI performs well. But in crises, when decisions can determine the survival of institutions or even the stability of the entire financial system, AI may make catastrophic errors or, worse, coordinate on destabilising strategies across institutions through common information channels. AI reliability is lowest precisely when it is needed most --- the AI wrong-way risk.

\subsection{ Procyclicality and Market Structure}\label{sec:channels_speed}

AI deployment in financial markets creates a powerful drive to more synchronised behaviour through two related channels: homogenisation of risk assessment methodologies and consolidation of market structure. Both channels reduce diversity in financial decision-making --- the first by promoting similar analytical approaches across institutions regardless of size and the second by concentrating market power among fewer, larger players. This twin reduction in diversity significantly alters how markets absorb and respond to shocks, inducing procyclicality with important implications for financial stability. Technically, this can happen when the correlation of institutions' signals rises, so the stabilising heterogeneity in equilibrium behaviour is reduced, increasing the risk of procyclical herd behaviour.

\subsubsection{Risk Monoculture}

The potential for synchronised behaviour is directly affected by two factors: the strength of strategic complementarities and the similarity in market participants' understanding of the world. When they have diverse views, their heterogeneous actions tend to absorb shocks --- some sell while others buy. Conversely, as information similarity increases, market participants become more likely to act in concert, amplifying both bubbles and crashes.

AI amplifies both drivers of synchronous behaviour. Beginning with information processing, while each AI system has its own neural network, the most successful architectures are likely to see widespread adoption. Financial institutions may independently converge on similar AI approaches because they prove effective, leading to what \citet{DanielssonMacraeUthemann2019} term ``risk monoculture'' --- a common view of risk that drives seemingly heterogeneous institutions to make similar decisions.

The use of common information sources further strengthens this convergence. Even when underlying network architectures differ across institutions, their training data often overlaps significantly. Major financial data vendors like Bloomberg, LSEG, and S\&P Global provide standardised data to multiple institutions. Financial regulations also create common data requirements, further homogenising the information landscape. This common substrate leads to similar interpretations of market conditions across different AI systems.

The speed advantage of AI may compound these effects by reducing the time available for diversity to emerge through human intervention.

\subsubsection{Market Concentration}

The cost structure of AI development further concentrates decision-making. As \citet{GambacortaShretti2025} observe, developing state-of-the-art AI systems requires significant investment in computing infrastructure, specialised talent and proprietary data --- resources available primarily to the largest financial institutions. This creates a tiered market structure where globally systemically important banks (GSIBs) can afford to develop proprietary AI engines, giving them significant competitive advantages in risk management and trading, while mid-tier institutions with legacy systems and traditional staff struggle to compete and must rely on off-the-shelf AI solutions from a limited number of vendors. Meanwhile, nimble fintech startups with modern technology stacks can leverage commodity AI tools creatively but lack the resources for fully customised solutions.

This stratification drives market concentration, potentially entrenching the dominance of GSIBs. Moreover, the competitive dynamics create a self-reinforcing cycle. Institutions that successfully deploy AI gain advantages that generate additional resources for AI investment, while laggards fall further behind. This winner-takes-most dynamic disproportionately benefits the largest institutions, further concentrating risk in systemically important entities.

\subsection{Speed}\label{sec:channels_speed}

The most immediate damage from financial crises stems from financial institutions' instinct to protect themselves during stress periods. When shocks occur, institutions must rapidly decide whether to run or to stay. This decision depends not only on fundamentals but critically on the action of others, with severe consequences for wrong decisions.

If an institution perceives that a shock is manageable, staying is optimal --- either maintaining positions or even purchasing assets sold in panic. However, if it concludes a crisis is imminent, running as quickly as possible becomes rational --- selling risky assets into falling markets and withdrawing liquidity. The first to run secures better prices; the last risks bankruptcy. This preemption dynamic creates powerful incentives for swift action once crisis signals emerge.

AI's speed advantage is crucial in this context. Our game-theoretic framework demonstrates that heterogeneous withdrawal capabilities substantially alter coordination thresholds. When some agents withdraw more effectively than others, the system becomes inherently more fragile. AI excels precisely at rapidly processing information and executing transactions, giving AI-powered institutions a significant edge during crises.

\subsubsection{AI vs. High-Frequency Trading}

While superficially similar to high-frequency trading (HFT), AI's impact on financial stability differs fundamentally. HFT algorithms typically operate with narrowly defined objectives within single securities or highly correlated asset pairs. As \citet{kirilenko2017flash} document, even these simple algorithms contributed to the 2010 Flash Crash when simultaneously withdrawing liquidity.

AI systems, however, can operate with complex objective functions spanning multiple asset classes and incorporating broader contextual information. Reinforcement learning dynamically adapts strategies based on changing market conditions, learns from interactions with other participants, and even coordinates implicitly on destabilising equilibria without explicit programming.

Conventional algorithmic trading generally improves liquidity under normal conditions \citep{Hendershott2011}, though high-frequency market making struggles when multiple assets experience simultaneous shocks \citep{brogaard2018high}. While sophisticated AI might better support liquidity in challenging situations through superior hedging strategies, its adaptive nature could reverse these benefits during stress periods by learning to exploit weaknesses in other algorithms or engaging in predatory trading \citep{BrunnermeierPedersen05}.

Importantly, while circuit breakers and trading halts have effectively contained HFT-driven volatility, such rule-based interventions may prove inadequate against AI systems that anticipate regulatory responses and develop sophisticated cross-market strategies to circumvent them. This adaptation capacity makes AI-driven synchronisation qualitatively different and potentially more systemically threatening than previous technologies.

\section{Optimal Policy Response}
 \label{sec:policy}

AI poses both opportunities and challenges for financial stability, and the authorities must respond to it. If they do so effectively, they can leverage AI to perform their mandate better. When they do not, they risk destabilising the financial system. We outline four critical areas arising from the model in Section \ref{sec:model} and the vulnerability channels in the previous Section, where financial authorities will have to respond to AI.

\subsection{Authority AI Systems}

To begin with, the authorities can benefit from developing their own AI capabilities to design regulations and evaluate intervention effectiveness. When AI first appeared on central banks' radar, many centred their response around divisions dealing with data, IT or innovation. However, given how AI can act as a powerful agent for systemic risk, financial stability should play a central role in AI policy. Model link: building in-house authority AI raises the effective cost of a hostile signal shift, reducing the probability that malicious information pushes beliefs past the run threshold.

For authority AI systems to be effective, the data they train on is particularly important. Current approaches to regulation rely on private sector firms providing voluminous PDF reports and database dumps, augmented by supervisors talking to private sector staff. Such a setup does not easily allow the authorities to gauge how the private sector would react to potential regulations and interventions. AI can significantly enhance this interaction by opening a new dimension in authority-private sector communication, making the supervisory process more robust and efficient. This requires developing a communication framework --- AI-to-AI links --- that allows an authority's AI engine to directly communicate with those of other authorities and the private sector.

AI-to-AI links allow real-time monitoring, allowing regulators to conduct performance benchmarking of private AI systems, simulate market stress, and evaluate their robustness and compliance with financial stability norms. The authority AI can query private sector AI and other agencies' AI on potential responses to particular interventions, aggregate those responses, and feed them back in an iterative process to find the ultimate impact of intervention strategies. This is similar to the Bank of England's System-wide Exploratory Scenario (SWES) exercise \citep{SWESboe}, but could be done considerably faster, involve more than two rounds and would be less taxing for the private sector.

\subsection{AI Monitoring Frameworks}

Traditional oversight mechanisms focus on capital, liquidity, and conduct but likely miss the unique systemic risks posed by AI. We propose developing comprehensive AI monitoring frameworks that track how AI is used across the financial system and assess potential coordination risks. To create an effective AI monitoring framework, financial authorities can establish a standardised reporting system --- an AI registry --- where financial institutions must document all AI systems deployed across their critical functions. 

\subsection{Crisis Facilities for AI-Driven Crises}

Central banks' standing facilities will likely become much more important as AI use proliferates. Central banks traditionally prefer discretionary facilities. However, such a deliberative approach may prove too slow in an AI-accelerated crisis. To address this temporal mismatch, automatic, pre-committed liquidity facilities that activate without human intervention would be of benefit. This approach has the additional benefit of potentially ruling out bad equilibria in our game-theoretic framework with public signals. If AI knows central banks will intervene if prices drop by a certain amount, they will not coordinate on strategies that are only profitable if prices drop more. Automatic standing facilities cap the price drop and raise the cut-off, discouraging the all-exit equilibrium that emerges when public signals are sufficiently precise.

\subsection{Public-Private Partnerships}

The optimal authority response to AI outlined above depends on authorities having access to AI engines that match the speed and complexity of private sector AI. It seems unlikely they will develop their in-house engines, as this would require considerable public investment and reorganisation. Instead, a more likely outcome is public-private partnerships that have already become common in financial regulation.

Another promising avenue is federated learning \citep{mcmahan2017communication}, a technique that enables institutions to train machine learning models locally on sensitive or proprietary datasets without exposing the underlying data. Instead of sharing raw data, only the trained model parameters or updates are transmitted to a central authority or aggregator. This approach enhances privacy and security while still allowing regulators to benefit from insights across multiple institutions. 
Sharing model architectures and network parameters limits the private speed advantage and dampens the procyclical rise in price impact that accelerates crises.

\section{Conclusion}
\label{sec:conclusion}

This paper examines the implications of artificial intelligence for financial stability through a game-theoretic framework. We identify three key mechanisms through which AI affects financial markets: superior information processing, common information sources and speed advantages. These mechanisms interact with endogenous system responses and strategic complementarities to create both stabilising and destabilising effects.

Our analysis suggests that AI will likely lower day-to-day volatility while increasing tail risk --- smoothing out short-term fluctuations at the expense of more extreme events. When faced with minor disturbances, AI can absorb shocks and stabilise markets. However, during genuine stress events, the same capabilities that dampen small fluctuations may amplify extreme movements, making crises faster and more intense than those we have experienced previously.

The financial authorities face significant challenges in adapting to this new landscape. Traditional regulatory frameworks designed for human decision-making timeframes may prove inadequate in markets where critical decisions unfold at machine speed. We propose specific policy responses, including authority AI systems, AI-to-AI communication frameworks, triggered facilities, public-private partnerships, and enhanced monitoring systems.

Ultimately, whether AI increases or decreases financial stability depends on how effectively both private market participants and public authorities harness its capabilities. If the financial authorities engage proactively with AI technology --- developing their capabilities, establishing direct communication channels with private sector AI, and designing appropriate automated response mechanisms --- they can leverage AI to enhance financial stability. If they do not, the likelihood of AI-amplified financial crises will increase. The key to financial stability in the age of AI lies not in resisting technological change but in ensuring that our regulatory frameworks evolve alongside it.

\clearpage

\bibliographystyle{chicago}

\clearpage

\begin{appendices}

\section{Solution of the Heterogeneous Global Game with AI and Human Agents}
\label{app:ai-crises-theory}

This appendix provides additional details on solving the heterogeneous global game presented in Section 3, particularly the derivation of the equilibrium threshold in the model with both human and AI agents. Our approach adapts the framework developed by \citet{SakovicsSteiner2012}.

\subsection{Belief Constraint in Heterogeneous Agent Models}

The key to solving the model is understanding what \citet{SakovicsSteiner2012} call the ``belief constraint,'' which governs the equilibrium relationship between the strategic beliefs of different types of agents and the distribution of the aggregate action.

In our model with two groups (humans $H$ and AI agents $A$), the belief constraint states that the weighted average of the strategic beliefs is the uniform distribution:

\begin{equation}
(1-\mu) A_H(a, \Delta) + \mu A_A(a, \Delta) = a
\end{equation}

where $A_g(a, \Delta)$ is the cumulative distribution function of the aggregate action as perceived by the threshold type of group $g$, $\Delta = (x_H^*$ - $x_A^*)/\sigma$ is a function of the threshold signals, and $\mu$ is the proportion of AI agents.

\subsection{Deriving the Threshold Equilibrium}

The equilibrium threshold $\theta^*$ is derived as follows:

\begin{enumerate}
\item Each threshold type satisfies an indifference condition. For a threshold type from group $g$, the expected payoff from investing must equal zero:
\begin{equation}
p_g \, b_g - c_g = 0
\end{equation}
where $p_g = \Pr(a \leq \theta^* | (x_g^*, g))$ is the probability the threshold type assigns to the absence of a crisis.

\item These probabilities can be expressed in terms of the strategic beliefs:
\begin{equation}
p_g = A_g(\theta^*, \Delta)
\end{equation}

\item From the indifference conditions, we get $p_g = c_g/b_g$ for each group.

\item Applying the belief constraint and substituting for $p_g$, we get:
\begin{equation}
\begin{aligned}
(1-\mu) A_H(\theta^*, \Delta) + \mu A_A(\theta^*, \Delta) &= \theta^* \\
(1-\mu) \frac{c_H}{b_H} + \mu \frac{c_A}{b_A} &= \theta^*
\end{aligned}
\end{equation}

\item In the special case where humans succeed with probability $p < 1$ when choosing to run while AI agents succeed with probability 1, the analysis from \citet{SakovicsSteiner2012} implies:
\begin{equation}
\theta^* = \frac{c}{b}[(1-\mu)p + \mu]
\end{equation}
where $c/b$ represents the common cost-benefit ratio when both groups have identical preferences.
\end{enumerate}

\subsection{Speed Advantage and Strategic Uncertainty}

The speed advantage of AI is modelled following the approach of \citet{SakovicsSteiner2012} for heterogeneous withdrawal capabilities. In this setup, the critical threshold $\theta^*$ increases with the measure $\mu$ of AI agents, making coordination on the run equilibrium more likely.

This happens because AI agents can execute transactions more quickly and efficiently than humans, creating a preemption motive. Consider the strategic uncertainty faced by threshold types:

\begin{itemize}
\item A human threshold type believes that AI agents are more likely to act before them, increasing their incentive to run.
\item An AI threshold type believes that it can act before humans, giving it a strategic advantage.
\end{itemize}

This asymmetry in strategic beliefs, while still satisfying the belief constraint, leads to a higher critical threshold than in a humans-only market (where $\theta^* = p \cdot c/b$).

The externalities created by AI's speed advantage can be calculated by differentiating the critical threshold with respect to $\mu$:

\begin{equation}
\frac{\partial \theta^*}{\partial \mu} = \frac{c}{b}(1-p) > 0
\end{equation}

This positive derivative indicates that increasing the proportion of AI agents makes the system more fragile.

\end{appendices}

\end{document}